\documentclass[sn-mathphys-ay]{sn-jnl}
\usepackage{natbib}
\usepackage[english]{babel}
\usepackage{mathptmx}
\usepackage{stmaryrd}
\usepackage{amssymb,amsmath}
\usepackage{float,graphicx,subfigure}
\usepackage{xcolor}

\newcommand{\Fig}[1]{Fig.~\ref{#1}}

\renewcommand{\vec}[1]{\mathbf{#1}}
\newcommand{\ld}{\,\mathrm{ld}\,}
\newcommand{\E}{\mathrm{e}}
\newcommand{\mng}[1]{\llbracket #1 \rrbracket}

\begin{document}


\title{Pragmatic information of aesthetic appraisal}

\author*[1]{\fnm{Peter } \spfx{beim} \sur{Graben}}\email{peter.beimgraben@b-tu.de}

\affil*[1]{\orgdiv{Bernstein Center for Computational Neuroscience}, \orgaddress{\city{Berlin}, \country{Germany}}}

\abstract{
A phenomenological model for aesthetic appraisal is proposed in terms of pragmatic information for a dynamic update semantics over belief states on an aesthetic appreciator. The model qualitatively correlates with aesthetic pleasure ratings in an experimental study on cadential effects in Western tonal music. Finally, related computational and neurodynamical accounts are discussed.
}


\date{\today}
\maketitle



\section{Introduction}
\label{sec:intro}
Aesthetics is, according to Alexander Gottlieb Baumgarten (1714 -- 1762), the science of beauty \citep{Baumgarten1750}. Putting it more scientifically, aesthetics is the science of critical judgement of the beautiful \citep{Nake74}. Therefore, aesthetics is also the science of critical judgement of the ugly as the opposite of the beautiful \citep{Stangneth19}. In these latter senses, aesthetics is part of aesthesiology as the science of perception in general, and of the reception of art in particular \citep{Consoli20, Tedesco24}, thus overlapping with psychology and the human cognitive neurosciences \citep{Fechner1876, FrascaroliLederEA24, PearceZaidelEA16}. For the judgment of an object or event as being beautiful, abstract features, such as regularity, symmetry or orderliness are often invoked \citep{Weyl80}. Formal principles such as the \emph{golden ratio} as one specific trait appear in all kinds of natural and artistic beauty, ranging from flowers to galaxies in nature and from ancient cultures to contemporary architecture, painting, sculpture, music, dance and drama \citep{Atlas03, Livio02, PowerShannon19}.

Also the pioneers of experimental psychology and psychophysics, Gustav Theodor Fechner (1801 -- 1887) and Wilhelm Wundt (1832 -- 1920) were acutely interested in aesthetic judgements. Fechner, e.g., in his \emph{Vorschule der \"Asthetik} (literally: \emph{Primary School on Aesthetics}), had conducted the probably very first experiments on empirical aesthetics: Subjects were asked to select the most ``beautiful'' and the ``ugliest'' quadrangle from a selection of ten simultaneously presented ones with equal areas but with different aspect ratios, ranging from squares ($1:1$) to rectangles with $2:5$, also including the golden ratio ($\varphi \approx 21:34 \approx 0.6176$) \citep[pp.~190]{Fechner1876}. His results indicated that the rectangle with the golden ratio was experienced as the most beautiful stimulus \citep[pp.~195]{Fechner1876}, although he conceded that the aesthetic value of the golden ratio could have been somewhat overestimated \citep[pp.~192]{Fechner1876}. Yet, later attempts to replicate Fechner's results have revealed some methodological flaws in his studies \citep[Ch.~7]{Livio02}.

Plotting the results of \citet[p.~195]{Fechner1876} graphically, entails a kind of bell-shaped, or ``inverse U-shaped'' response curve that already appeared in Wundt's \emph{Grund\-züge der Physiologischen Psychologie} (\emph{Principles of Physiological Psychology}) as emotional strength depending on stimulus intensity \citep[pp.~432]{Wundt1874}. This characteristic curve (\citet[Fig.~97]{Wundt1874}; also compare with \Fig{fig:entropy} below for illustration)
was dubbed the ``Wundt curve'' of hedonic value by \citet{Berlyne70}, where
\begin{quote}
  [\dots] the abscissa is taken to represent ``arousal potential,'' a term meant to cover all the stimulus properties that tend to raise arousal, including novelty and complexity. Positive hedonic value reaches a maximum with moderate arousal potential (stimulation producing a moderate arousal increment) and then, as arousal potential increases further, hedonic value takes on lower and lower positive values and finally becomes negative. \citep[p.~284]{Berlyne70}
\end{quote}
Here, maximal ``hedonic value'' corresponds with the aesthetic appreciation of the beautiful.

The Wundt curve of aesthetic appraisal has been confirmed more or less in numerous studies on empirical aesthetics. To mention only a few of them: \citet{MullenArnold76} presented pairs of rhythmic musical sequences to their subjects. These were asked to choose the most preferred or most interesting sequences. As one result they reported that the beginning of a response curve displays an inverted U-shape in dependence on redundancy in rhythmic sequences. In a more recent study, \citet{MarinLeder13} proposed a cross-domain approach, acknowledging the multidimensional nature of complexity. In four experiments, they employed pictures of affective scenes, representational paintings, and music excerpts. Stimuli were pre-selected to vary in emotional content and complexity. Ratings of familiarity, complexity, pleasantness and arousal were obtained for each set of stimuli. Subsequent statistical analyses revealed positive correlations between subjective complexity and arousal. Later, \citet{GucluturkJacobsLier16} collected ratings of liking and perceived complexity for digitally generated images. Additionally, they also calculated complexity measures for each image. As a result, they reported an inverted U-curve relationship between aesthetic appreciation and stimulus complexity. In a large corpus study, \citet{SigakiPercRibeiro18} and \citet{Perc20} conducted quantitative analyses of almost 140,000 paintings, covering nearly a millennium of art history. Based on the local spatial patterns in the images, they estimated entropy and complexity measures of each painting. These measures represented the amount of visual order of artworks, locally reflecting qualitative categories proposed by art historians. Interestingly, the dependence of statistical complexity on entropy exhibited another inverted U-curve relationship. Finally, \citet{AgresHerremansEA17} investigated the connection between harmonic musical structure and aesthetic appraisal. In a psychological rating experiment listeners assessed their enjoyment for computer-generated anthems with varying levels of harmonic repetition and complexity. These ratings were further explored using computational approaches, based on information theory and on musicological tonal tension measures \citep{YouSunYang23}. Their results also provided evidence for a Wundt-type relationship between complexity and aesthetic enjoyment.

Clearly, judgements of the beautiful and the reception of art works are complex neurodynamical and mental processes that require multivariate and multidimensional methodology \citep{Consoli20}, which was substantially recognized in the experiments by \citet{CheungHarrisonEA19, CheungHarrisonEA24} on the appreciation of musical beauty. Specifically, \citet{CheungHarrisonEA19} demonstrated that pleasure varies nonlinearly as a function of two independent variables: context uncertainty and stimulus surprisal. Taking Western tonal harmony as a model of musical structure, they trained machine-learning models to mathematically quantify the uncertainty and surprise of about 80,000 chords in pop songs. They reported that cadential chords were experienced as most pleasurable in two distinguished cases: On the one hand for contexts with low uncertainty but highly surprising closure. On the other hand for highly uncertain contexts, but less surprising closure. In an additional neuroimaging study, \citet{CheungHarrisonEA19} reported that activity in the amygdala, hippocampus, and auditory cortex reflected the observed interaction, while the nucleus accumbens only accounts for uncertainty. It is the aim of the present study, to discuss those results, partly reproduced in \Fig{fig:praginfo}(a) below, in the light of a phenomenological model of aesthetic reception.

As already mentioned, many of the studies reviewed above, employed methods from information theory \citep{Moles66, Nake74, Nake12, ShannonWeaver49, Volz88}, complexity science \citep{BadiiPoliti97, Gernert06}, and machine learning \citep{Pearce18, RussellNorvig10, Schmidhuber10a} which are also suitable for a deeper theoretical understanding and philosophical reflection of aesthetic judgements. An early account was given by Immanuel Kant (1724 -- 1804) in his \emph{Kritik der Urteilskraft} (\emph{Critique of the Power of Judgement}), where he wrote:
\begin{quote}
    [\dots] the judgement of taste must rest on a mere sensation of the reciprocal activity of the Imagination in its freedom and the Understanding with its conformity to law. It must therefore rest on a feeling, which makes us judge the object
    by the purposiveness of the representation (by which an object is given) in respect of the furtherance of the cognitive faculty in its free play. \citep[p.~161, §35]{Kant14}
\end{quote}
For Kant, a ``judgement of taste'' (i.e., an aesthetic judgement) involves two mental faculties: ``Understanding'' as the ``faculty of rules'' \citep[B171]{Kant99a} and ``Imagination'' as the ``faculty of intuitions'' \citep[p.~161, §35]{Kant14} that are engaged in a free, harmonious play, enjoying the recipient's mind upon the beautiful. Thus, the faculty of ``Understanding'' always tries to classify a beautiful object or event according to some rules of regularity, principles of symmetry, or laws of orderliness. Contrastingly, the faculty of ``Imagination'' permanently attempts to generate novel aspects, creative insights and fantastic associations of the given object or event \citep{Graben24b}.

These characteristics of aesthetic judgements were also emphasized in the endeavor of information-theoretical aesthetics \citep{Moles66, Nake74, Nake12, Volz88}. According to \citet{Moles66}, complete regularity and order are experienced as \emph{banality}, while irregular creativity and \emph{originality} appear as randomness---both extremes not exhibiting any hedonic value. Applied to music reception, \citet{Meyer56} likewise observed that if
\begin{quote}
   the stimuli comprising the series cannot be perceived as being similar in any respect whatsoever, then they will fail to cohere, to form a group or unit, and will be perceived as separate, isolated, and discrete sounds, ``signifying nothing.'' Since contrast and comparison can exist only where there is similarity or equality of some sort, the mental impression created by such a series will be one of dispersion, not disparity; of diffusion, not divergence; of novelty, not variety. [\dots] Complete similarity, proximity, and equality of stimulation, on the other hand, will create an undifferentiated homogeneity out of which no relationships can arise because there are no separable, individual identities to be contrasted, compared, or otherwise related. There will be coexistence and constancy, but not connection; uniformity and union, but not unity. In short, both total segregation and total uniformity will produce sensation, but neither will be apprehended as pattern or shape. \citep[p.~158]{Meyer56}
\end{quote}

Here again, the dichotomy between ``total segregation and total uniformity'' corresponds to the distinction between
``novelty'', or ``originality'' \citep{Moles66}, on the one hand, and ``similarity'', or ``banality'' \citep{Moles66}, on the other hand. It is the aim of the present study to formalize this dichotomy in terms of information theory, thereby utilizing the resulting phenomenological model of \emph{pragmatic information} \citep{Gernert06, Graben06, KornwachsLucadou82, WeizsackerWeizsacker72} as a correlate of aesthetic appraisal \citep{MenninghausWagnerEA19, Blutner24b}.


\section{Pragmatic Information}
\label{sec:pi}
In his own contribution to the foundation of information theory, Weaver described three levels of the communication problem ``by which one mind may affect another'' \citep[p. 3]{ShannonWeaver49}, namely
\begin{quote}
\begin{itemize}
  \item Level A. How accurately can the symbols of communication be transmitted? (The technical
problem.)
  \item Level B. How precisely do the transmitted symbols convey the desired meaning? (The
semantic problem.)
  \item Level C. How effectively does the received meaning affect conduct in the desired way?
(The effectiveness problem.) \citep[p.~4]{ShannonWeaver49}.
\end{itemize}
\end{quote}
These levels correspond roughly with the dimensions of semiotics \citep{Morris38}. The ``technical problem'' with noisy communication channels can be addressed by redundant codes, introducing correlations between the symbols of the messages, i.e.,~by \emph{syntax}. The ``semantic problem'' refers to correlations between the transmitted symbols and their intended meanings (\emph{semantics}). Finally, the ``effectiveness problem'' relates to \emph{pragmatics}, namely the correlations between the symbols and their impact upon the message users.

Shannon's information measure $H$ which is formally related to Boltzmann's entropy in statistical physics and thermodynamics only accounts for the ``technical problem'' of efficient communication \citep{ShannonWeaver49}. A tentative measure of semantic information, or informativity $I$, was proposed by \citet{Bar-HillelCarnap53}. Finally, \citet{WeizsackerWeizsacker72} suggested three desiderata to be fulfilled by any reasonable measure of pragmatic information $S$:
\begin{itemize}
  \item (\emph{i}) Pragmatic information $S$ assesses the impact of a message upon the mental states of a cognitive agent.
  \item (\emph{ii}) Pragmatic information $S$ should vanish in the limits of novelty (originality) $N$ and confirmation (banality) $C$.
  \item (\emph{iii}) Novelty and confirmation presumably require a non-classical but rather quantum-like description in terms of incompatible mental operators.
\end{itemize}
Obviously, condition (\emph{ii}) could be easily satisfied by defining pragmatic information as a product of suitable measures for novelty and confirmation: $S = N \times C$, such that $S = 0$ for either $N = 0$ or $C = 0$. This idea was originally suggested by \citet{KornwachsLucadou82}.

In Shannon's information theory \citep{ShannonWeaver49}, a discrete information source is given by an $n$-dimensional probability space $(\mathcal{X}, \vec{p})$ endowed with a probability distribution $\vec{p} \in [0, 1]^n$, such that $0 \le p_i \le 1$ is the probability of the atomic event $E_i \in \mathcal{X}$. The discrete set $\mathcal{X}$ is considered as a \emph{repertoire} with $p_i$ being the probability that the event $E_i$ was selected \citep{Nake74, Nake12}.

The information content, or \emph{informativity} \citep{Bar-HillelCarnap53}, of an event $E_i$ is defined as
\begin{equation}\label{eq:info}
  I(p_i) = - \ld p_i \:.
\end{equation}
The base of the logarithm is usually chosen as two, such that  $\ld \equiv \log_2$ yields information measured in ``bits''. Equation \eqref{eq:info} indicates that the smaller the probability $p_i$, the larger the informativity $I(p_i)$, which is therefore a measure of the \emph{surprise} of the event $E_i$ \citep{AgresAbdallahPearce18, Hale16}. Clearly, informativity $I$ can also be regarded as a measure of novelty or originality $N$ \citep{WeizsackerWeizsacker72}.

The averaged information over the entire distribution entails the Shannon \emph{entropy}
\begin{equation}\label{eq:entropy}
    H(\vec{p}) = \sum_i p_i \, I(p_i) = - \sum_i p_i \, \ld p_i \:,
\end{equation}
where the natural convention $p_i \ld p_i = 0$ if $p_i = 0$ applies \citep{ShannonWeaver49}. Note that the contribution of a single event $E_i \in \mathcal{X}$ to the entropy of the probability space,
\begin{equation}\label{eq:conspic}
    S(p_i) = p_i \, I(p_i) = - p_i \, \ld p_i = N(p_i) \times C(p_i) \:,
\end{equation}
could be regarded as a measure of pragmatic information of $E_i$ in the sense of \citet{KornwachsLucadou82}, for $N(p_i) = I(p_i)$ assesses novelty, while $C(p_i) = p_i$ obviously reflects confirmation. This quantity has been introduced as the \emph{conspicuity} (\emph{Auff\"alligkeit}) of the event $E_i$ in the framework of information-theoretic aesthetics \citep{Nake74, Nake12, Volz88}. Further notice that the function \eqref{eq:conspic} assumes its local maximum at site $p_i^* = 1/\E \approx 0.3679$ which is surprisingly close to the complement of the golden ratio $1 - \varphi \approx 0.3824$ \citep{Nake74, Volz88}.

In order to fulfill the three desiderata of \citet{WeizsackerWeizsacker72} above, an appropriate framework for music reception had already been proposed by \citet{Meyer56}, that can be straightforwardly applied to aesthetic appreciation in general:
\begin{quote}
  We have stated that styles in music are basically complex systems of probability relationships [\dots] \citep[p.~54]{Meyer56}

  The listener brings to music not only specifically musical experiences, associations, and dispositions but also important beliefs as to the nature and significance of aesthetic experience in general and the expected musical experience in particular. \citep[p.~73]{Meyer56}

  Revision of opinion, stressed earlier in the discussion of probability, is also important in the perception of form. Here, too, the listener often finds it necessary to revise his opinions of the significance of what has passed and his expectations
  of what is still to come in the light of an unexpected present. \citep[p.~60]{Meyer56}

  But even more important than designative meaning is what we have called embodied meaning. From this point of view what a musical stimulus or a series of stimuli indicate and point to are not extramusical concepts and objects but other musical events which are about to happen. That is, one musical event (be it a tone, a phrase, or a whole section) has meaning because it points to and makes us expect another musical event. \citep[p.~35]{Meyer56}
\end{quote}
This small selection of quotes motivates the following definitions.

Let $\mathcal{A}$ be a cognitive agent, here regarded as an aesthetic \emph{appreciator} \citep{Nake74} with mental belief space $(\mathcal{X}, \vec{p})$. Thus, a mental state of $\mathcal{A}$ is a probability distribution $\vec{p}$ over a repertoire $\mathcal{X}$, or briefly, a \emph{belief state} over $\mathcal{X}$ in the sense of either partially observable decision models in artificial intelligence research \citep{RussellNorvig10}, or in the sense of Bayesian dynamic update semantics \citep{BenthemGerbrandyKooi09, Gardenfors88, Graben06, Graben14a}.

Next, consider a natural or artistic object or event $A$, such as a phrase or a verse of a poem, a view upon a building, a perspective toward a sculpture, a scene in a drama, or a motive or a phrase in music. When $A$ can be represented as a mental operator $\vec{A}$ acting on $(\mathcal{X}, \vec{p})$ by means of a suitable functor $\Phi: A \mapsto \vec{A}$, a mapping from an \emph{a priori} belief state $\vec{q}$ (i.e., a distribution over the repertoire $\mathcal{X}$) to an \emph{a posteriori} belief state $\vec{p}$ (i.e., another or the same distribution), defined by the action
\begin{equation}\label{eq:update}
    \vec{p} = \vec{A}\, \vec{q}
\end{equation}
of $\vec{A}$, entails the dynamic meaning $\mng{A}$ of the aesthetic object or event $A$, hence
\begin{equation}\label{eq:meaning}
    \mng{A} = \vec{A}: \vec{q} \mapsto \vec{p} \:.
\end{equation}
Note that $\vec{A}$ could be in general a nonlinear mapping from prior to posterior. In case $\vec{A}$ is a linear operator, it becomes a Markovian stochastic transition matrix as often employed for statistical learning approaches to aesthetic enculturation \citep{AgresAbdallahPearce18, Pearce18}. Further note, that, if belief states could be expressed as squared projections on Hilbert space subspaces, the present approach fits into the framework of quantum-inspired cognitive systems \citep{HuberRomerEA22}. Hence, the third desideratum of \citet{WeizsackerWeizsacker72} would be naturally fulfilled.

Now, the \emph{information gain} for a single element $E_i$ of the repertoire $\mathcal{X}$ \citep{Kharkevich60} that is provided by the application of the aesthetic operator $\vec{A}$ upon a belief state $\vec{q}$, can be expressed as the difference between the information contents of the prior and its updated posterior,
\begin{equation}\label{eq:infogain}
    \Delta I(p_i) = I(q_i) - I(p_i) = - \ld q_i - (- \ld p_i) = \ld \frac{p_i}{q_i} \:.
\end{equation}

Averaging $\Delta I(p_i)$ over all events $E_i \in \mathcal{X}$ with respect to the updated distribution $\vec{p}$ yields the so-called \emph{Kullback-Leibler divergence} as the average information gain \citep{KullbackLeibler51} between the distributions $\vec{p}, \vec{q}$,
\begin{equation}\label{eq:KL}
    K(\vec{p} \| \vec{q}) = \sum_i p_i \Delta I(p_i) = \sum_i p_i \ld \frac{p_i}{q_i} \:.
\end{equation}

Under the special constraint that $\vec{A}$ satisfies every desideratum of \citet{WeizsackerWeizsacker72}, listed above, the Kullback-Leibler divergence
\begin{equation}\label{eq:pi}
    S(\vec{p} \| \vec{q}) = K(\vec{A} \vec{q} \| \vec{q})
\end{equation}
is called \emph{pragmatic information} of the aesthetic operator $\vec{A}$ in belief state $\vec{q}$ for the dynamic updating \eqref{eq:update}: $\vec{p} = \vec{A}\, \vec{q}$ \citep{Graben06}. Note that \citet{AgresAbdallahPearce18} refer to Kullback-Leibler divergence as to \emph{predictive information} in their own statistical learning approach, whereas \citet{Skyrms10} used Kullback-Leibler divergence for the assessment of signals in pragmatic signaling games.


\section{Modeling Aesthetic Appraisal}
\label{sec:maa}
As already mentioned in the Introduction, the present study aims at a particular application of the phenomenological model for aesthetic pragmatic information, outlined in the previous section, to the experimental findings of \citet{CheungHarrisonEA19}. In their first experiment, 39 healthy adults had to listen to 1,039 chords in 30 chord progressions that were selected from 745 pop songs listed in the US Billboard ``‘Hot 100'' chart between 1958 and 1991, after removing melody and rhythm and after transposition to C major mode. Subjects were instructed to continuously rate the pleasantness of each chord using a mechanical slider during stimulus exposure. To obtain the independent variables, an unsupervised statistical-learning model \citep{Pearce18} was trained to assess the statistical regularities of over 80,000 chord progressions from the same music corpus. For each chord presented in the rating experiment, its uncertainty $H$ was estimated as conditional Shannon entropy of the model. Similarly, the surprise $I$ was determined for every chord through the model. Figure \Fig{fig:praginfo}(a) shows the results of \citet{CheungHarrisonEA19}. It displays standardized pleasure ratings $P(I \| H)$, namely the aesthetic appreciation of music, in dependence of two variables: context uncertainty, $H(q)$, and stimulus surprise, $I(p)$.

In order to develop a mathematical model for the data plotted in \Fig{fig:praginfo}(a), one has to ascertain functional relationships between the experimental variables of \citet{CheungHarrisonEA19}. For the sake of simplicity, one considers a binary repertoire $\mathcal{X} = \{ E, F \}$ without any statistical correlations between successive events. Then, the prior $\vec{q} = (q, 1 - q)$ as well as the posterior $\vec{p} = (p, 1 - p)$ are parameterized by only one real number each: $q \in [0, 1]$ for the \emph{a priori} distribution, and $p \in [0, 1]$ for the dynamically updated \emph{a posteriori} distribution, with $\vec{p} = \vec{A}\, \vec{q}$ under the aesthetic operator $\vec{A}$.

\begin{figure}[H]
    \includegraphics[width=0.9\textwidth]{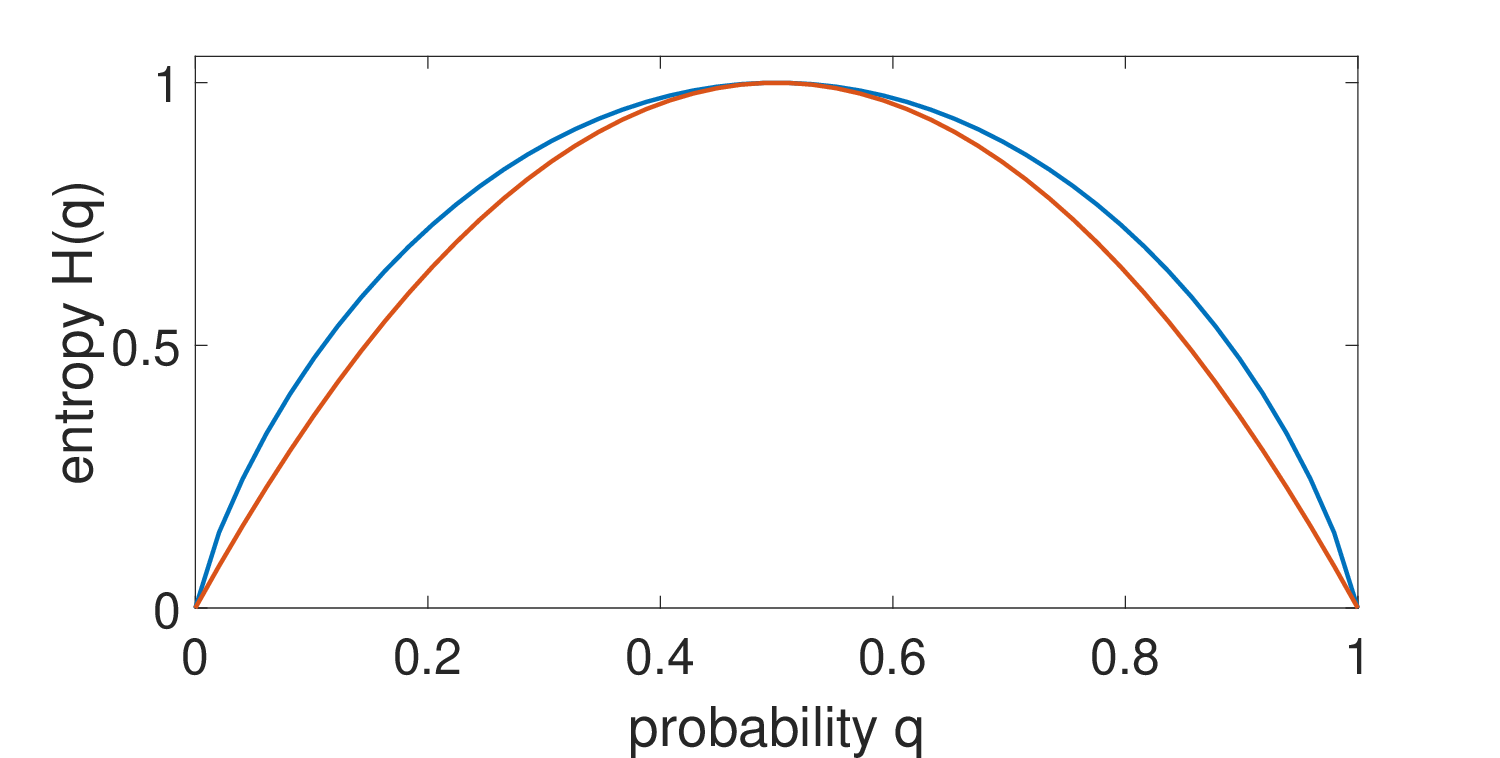}
    \caption{\label{fig:entropy}
    Comparison of Fermi-Dirac entropy $H(q)$ with logistic map $L(q)$ against probability parameter $q$ of a binary distribution. Blue: Fermi-Dirac entropy \eqref{eq:fdent}; red: logistic map \eqref{eq:logi}.
}
\end{figure}

Under those assumptions, the Shannon entropy \eqref{eq:entropy} of the prior,
\begin{equation}\label{eq:fdent}
  H(q) = - q \ld q - (1 - q) \ld (1 - q) \:,
\end{equation}
which is known as Fermi-Dirac entropy in this particular form in statistical physics as it is maximized by the Fermi-Dirac distribution in the respective variational calculus \citep{Kapur70, BorweinReichSabach11}, becomes a suitable measure of context uncertainty for the study of \citet{CheungHarrisonEA19}. Figure \ref{fig:entropy} displays the Fermi-Dirac entropy as a blue curve.

By contrast, the informativity \eqref{eq:info}
\begin{equation}\label{eq:info2}
  I(p) = - \ld p
\end{equation}
may be appropriately attributed to the surprise of the aesthetically closing stimulus $\vec{A}$ in the experiment of \citet{CheungHarrisonEA19}. Finally, the Kullback-Leibler divergence \eqref{eq:KL}, interpreted as pragmatic information \eqref{eq:pi},
\begin{equation}\label{eq:pi2}
  S(p \| q) =  p \ld \frac{p}{q} + (1 - p) \ld \frac{1 - p}{1 - q}
\end{equation}
should be correlated with aesthetic enjoyment $P(I \| H)$ \citep{CheungHarrisonEA19} for all possible combinations of $p, q \in [0, 1]$.

To further proceed, one has to substitute all occurrences of $p, q$ in \eqref{eq:pi2} by $H(q)$ and $I(p)$, inverting the functions  \eqref{eq:fdent} and \eqref{eq:info2}, respectively. For the latter, this is straightforwardly achieved through
\begin{equation}\label{eq:invinfo}
  p = 2^{-I(p)} \:.
\end{equation}
However, the Fermi-Dirac entropy \eqref{eq:fdent} can be firstly approximated by the logistic map
\begin{equation}\label{eq:logi}
  L(q) = 4 q (1 - q)
\end{equation}
plotted as the red curve in \Fig{fig:entropy} above. The logistic map is well-known from dynamical system theory \citep{EckmannRuelle85}.\footnote{
    I thank Reinhard Blutner for suggesting this approximation in a recent personal communication.
}
Its inverse function is easily obtained through two branches
\begin{equation}\label{eq:loginv}
  q_\pm = \frac{1}{2} \left( 1 \pm \sqrt{1 - L(q)}\right)
\end{equation}
with a lower branch $q_-$ and an upper branch $q_+$.

Inserting \eqref{eq:loginv} and \eqref{eq:invinfo} into \eqref{eq:pi2}, by exploiting the approximation $H(q) \approx L(q)$, yields the expression
\begin{equation}\label{eq:pi3}
  S(I(p) \| H(q)) =  2^{-I(p)} \ld \frac{2^{1 - I(p)}}{1 \pm \sqrt{1 - H(q)}} +
        (1 - 2^{-I(p)}) \ld \frac{2 - 2^{ 1 - I(p)}}{1 \mp \sqrt{1 - H(q)}}
\end{equation}
as a phenomenological model for aesthetic appraisal. Note that the function $S(I(p) \| H(q))$ is divergent for $H \to 0$, such that it is not well-defined along the ordinate.


\section{Results}
\label{sec:res}
\begin{figure}[H]
\centering
\subfigure[]{\includegraphics[width=0.9\textwidth]{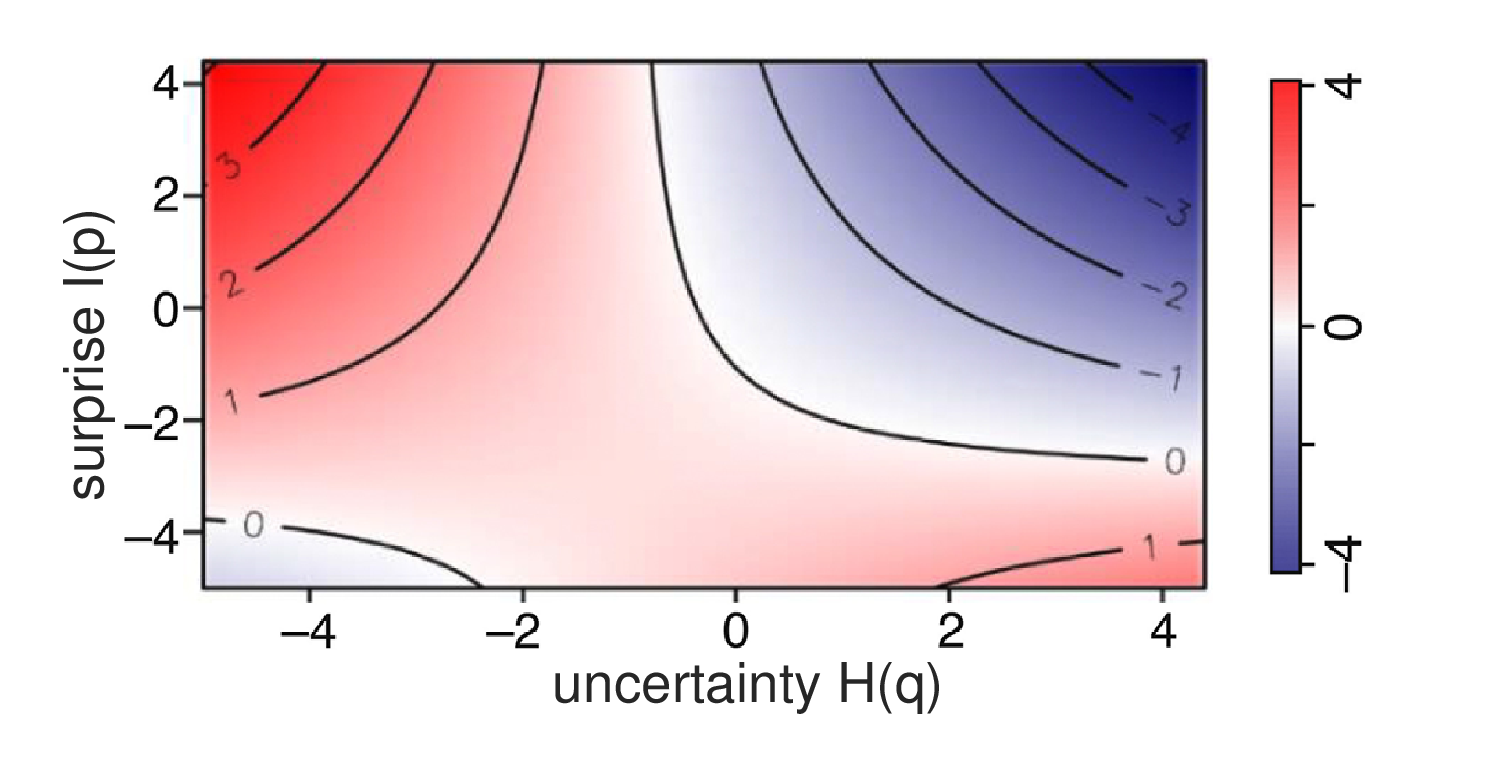}} \qquad
\subfigure[]{\includegraphics[width=0.9\textwidth]{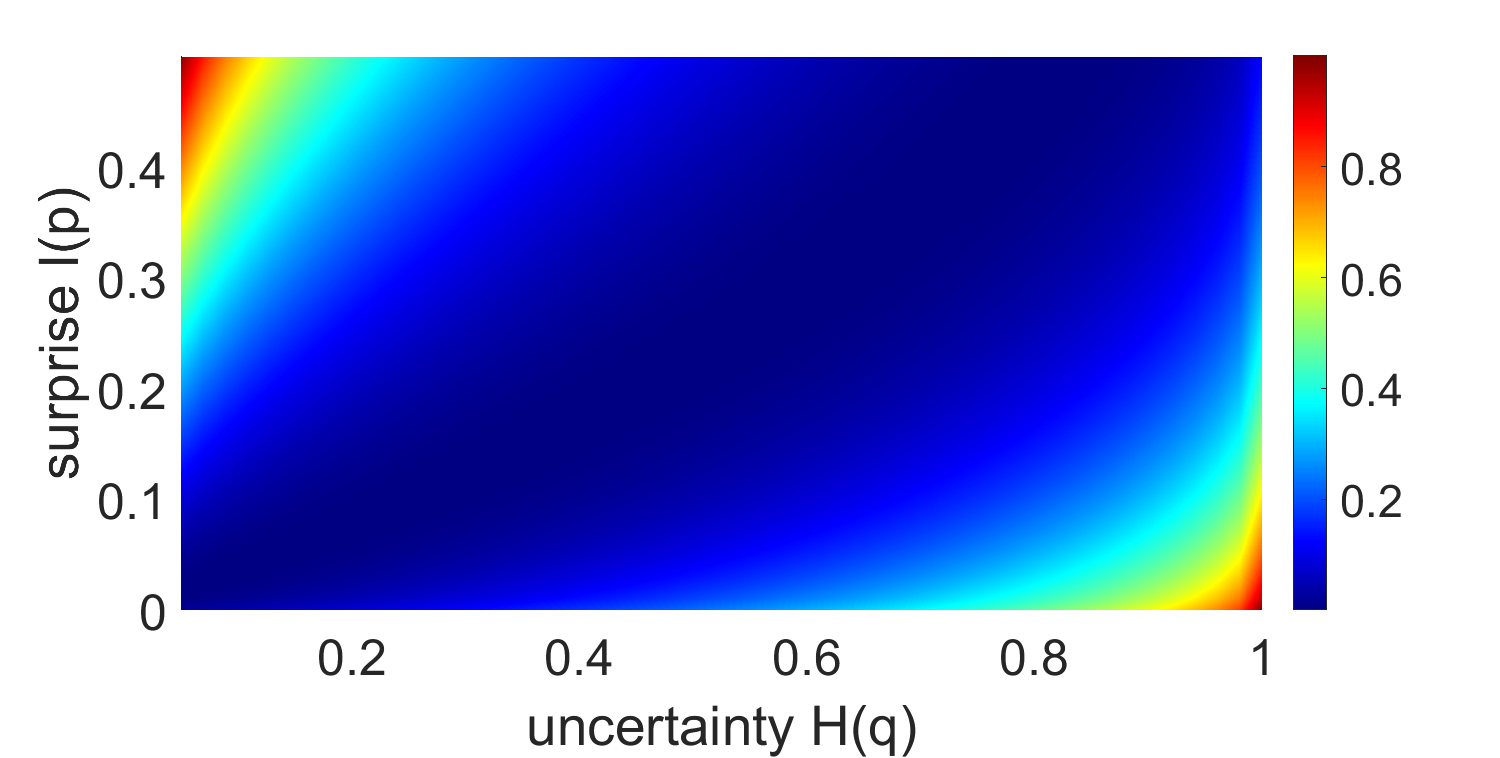}}
\caption{\label{fig:praginfo}
    Comparison of aesthetic pleasure rating $P(I \| H)$ from the experimental study of \citet{CheungHarrisonEA19} with pragmatic information model $S(I(p) \| H(q))$. (a) Normalized $z$-scores of aesthetic appreciation in dependence of context uncertainty $H(q)$ (abscissa) and stimulus surprise $I(p)$ (ordinate). Reprinted after slight modifications from \citet[Fig.~2(b)]{CheungHarrisonEA19} with permission from Elsevier. (b) Pragmatic information \eqref{eq:pi3} for probabilistic context updating dynamics \eqref{eq:update} under a cadential operator $\vec{A}$.
}
\end{figure}

The results of the phenomenological account for pragmatic information of aesthetic appraisal are shown in \Fig{fig:praginfo}. Here, \Fig{fig:praginfo}(a), reproduced from \citet{CheungHarrisonEA19} with permission of Elsevier, depicts the standardized aesthetic pleasure measure $P(I \| H)$ gathered in the musical rating experiment of \citet{CheungHarrisonEA19} on the one hand. The color coding reflects the emotional intensity of the subjects' pleasure rating, ranging from $P(I \| H) = -4$ (dark blue)  to $P(I \| H) = +4$ (dark red). The independent variables are context uncertainty $H(q)$ of the \emph{a priori} belief state $\vec{q}$ plotted along the abscissa, and stimulus surprise (ordinate), measured as informativity $I(p)$ of the \emph{a posteriori} belief state $\vec{p}$ under aesthetic dynamic updating, $\vec{p} = \vec{A} \vec{q}$, for a cadential chord $\vec{A}$.

On the other hand, \Fig{fig:praginfo}(b) presents the Kullback-Leibler divergence \eqref{eq:pi3} computed for the upper branch $q_+$ of the Fermi-Dirac entropy $H(q)$ \eqref{eq:fdent} plotted as a model of context uncertainty along the abscissa. The ordinate depicts the informativity $I(p)$ of the \emph{a posteriori} belief state $\vec{p}$ modeled through \eqref{eq:info2}. Again, large positive values are indicated by bright red, while very small values are represented by dark blue, here. Because of the divergence of $S(I(p) \| H(q))$ for $H \to 0$, only an interval $[\varepsilon, 1]$ is shown for the abscissa with small cut-off $\varepsilon = 0.05$. The ordinate is plotted for $[0, 0.5]$.

Obviously, there is a remarkable qualitative correlation between the experimental rating data in \Fig{fig:praginfo}(a) and the phenomenologically modeled pragmatic information $S(I(p) \| H(q))$ in \Fig{fig:praginfo}(b). Pragmatic information is high for quite predictable stimuli (low $I(p)$) in very uncertain contexts (high $H(q)$), and also for highly surprising stimuli (high $I(p)$) in rather certain contexts (low $H(q)$), thereby reflecting the aesthetic appeal of the cadential closures $\vec{A}$ in Western tonal music. By contrast, aesthetic pleasure and also pragmatic information are low for expected stimuli (low $I(p)$) appearing in certain contexts (low $H(q)$), characterizing boredom and banality, as well as for surprising stimuli (high $I(p)$) in very uncertain contexts (high $H(q)$), reflecting randomness or originality \citep{Moles66, Nake74, Nake12}. This finding confirms the second desideratum of \citet{WeizsackerWeizsacker72}. However, both quantities differ in the respect, that the experimental data exhibit a saddle in parameter space which is not present for pragmatic information, due to its non-negativity constraint.

\begin{figure}[H]
    \includegraphics[width=0.9\textwidth]{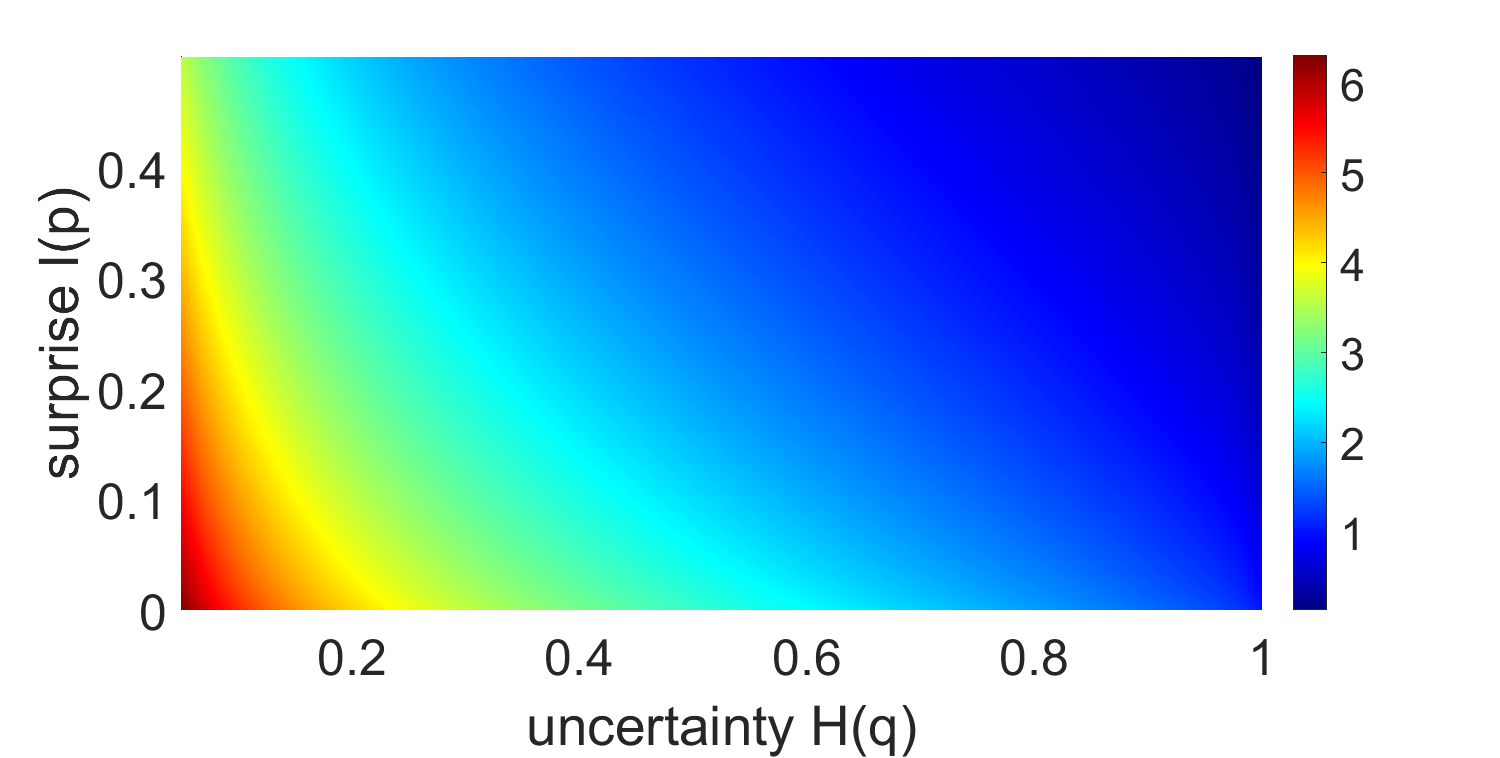}
    \caption{\label{fig:negbranch} Kullback-Leibler divergence for the lower branch of the probabilistic context updating dynamics \eqref{eq:update}, depending on context uncertainty $H(q)$ (abscissa) and stimulus surprise $I(p)$ (ordinate).
}
\end{figure}

For the sake of completeness, \Fig{fig:negbranch} shows additionally the Kullback-Leibler divergence \eqref{eq:pi3} computed for the lower branch $q_-$ of the Fermi-Dirac entropy \eqref{eq:fdent}.

Interestingly, this picture cannot consistently be interpreted as pragmatic information because of the prevailing divergence of $S(I(p) \| H(q))$ for $H \to 0$ that is even present for the interval cut-off $\varepsilon = 0.05$ used in \Fig{fig:praginfo}(b). This behavior indicates that Kullback-Leibler divergence for the lower Fermi-Dirac branch is large for boring stimuli appearing in banal contexts, thus violating the second desideratum of \citet{WeizsackerWeizsacker72} for pragmatic information. It is therefore also not a viable correlate of aesthetic appreciation.


\section{Discussion}
\label{sec:disc}
In their original proposal toward a theory of pragmatic information, \citet{WeizsackerWeizsacker72} formulated three desiderata that could be naturally fulfilled in the framework of dynamic update semantics \citep{Graben06}. First, pragmatic information ascertains the impact of a message or event upon the mental state space of a cognitive agent. In dynamic semantics and machine learning, mental states of a cognitive agent are usually modeled as belief states, namely probability distributions over cognitive repertoires that become updated by novel evidence provided by the reception of messages or events \citep{BenthemGerbrandyKooi09, Gardenfors88, Graben06, Graben14a, RussellNorvig10}. Second, pragmatic information vanishes in the limits of novelty and confirmation. In dynamic semantics a message or event exhibits novelty if it cannot be understood or even processed by a cognitive agent, whereas confirmation can be expressed by means of logical consequence. Computing the Kullback-Leibler divergence for both special kinds of dynamic updating entails naught. Thus, \citet{Graben06} suggested this quantity as a suitable measure of pragmatic information. Third, pragmatic information requires a quantum-like treatment of incompatible operators. In dynamic semantics, belief revision, default reasoning, and anaphora present paradigmatic examples of non-commutative and hence incompatible updating processes \citep{Graben06, Graben14a}. Moreover, a description of belief states in terms of Hilbert space vectors leads straightforwardly to a quantum-like formalism \citep{HuberRomerEA22}. Related approaches of pragmatic information have been used by \citet{AtmanspacherScheingraber90, Gernert06} and, more recently, by \citet{DavisSchubelerKozma24}.

The present study has demonstrated that the framework of pragmatic information in dynamic semantics can be immediately applied to aesthetics as the science of beauty \citep{Baumgarten1750}. Under the assumption that aesthetic objects or events act as quantum-like operators upon the belief states of a cognitive agent \citep{Meyer56}, called the \emph{appreciator} in the present context \citep{Nake74}, hedonic  value is minimized in cases of plain banality and overburdening originality \citep{Moles66, Nake74, Nake12, Volz88}. Applied to a recent study on the appreciation of closure effects (cadences) in Western tonal music \citep{CheungHarrisonEA19}, the suggested measure of pragmatic information provides a phenomenological model of the experimental results. Vanishing in the cases of banality and originality, pragmatic information qualitatively correlates with high aesthetic pleasure for predictable closures in rather uncertain contexts and also for surprising closures in more certain contexts.

The unpretentious phenomenological model of aesthetic appraisal presented here, is based on a binary repertoire $\mathcal{X} \equiv \mathcal{X}_0$ of maximal coarse grain that produces a stochastic Bernoulli process of uncorrelated events. In order to discuss computational models of aesthetic appreciation as well, one has to consider a hierarchy of repertoires $\mathcal{X}_k$ ($k \in \mathbb{N}_0$), where $\mathcal{X}_{k + 1}$ is a refinement of $\mathcal{X}_k$. \Citet{Nake74, Nake12} described aesthetic appreciation as an iterative algorithm that converges toward some optimal $\mathcal{X}_n$ where the Shannon entropy of the repertoire $\mathcal{X}_n$ has dropped below a given memory capacity, while that of its immediate refinement $\mathcal{X}_{n+1}$ slightly exceeds that threshold.

A related, though quite different account, was suggested by \citet{Mizraji23}. He illustrated the effect of aesthetic emotions  by means of photomosaics \citep{Silvers96}, where a repertoire of a certain grain $\mathcal{X}_n$ is provided by the mosaic tiles comprising some picture on a much larger scale. \Citet{Graben24b} has further augmented \citeauthor{Mizraji23}'s \citeyearpar{Mizraji23} argument by drawing an analogy to the \emph{Beauty of Fractals} \citep{PeitgenRichter86}. This ana\-logy allows for an interpretation of Kant's philosophical aesthetics, mentioned in the Introduction \citep{Kant14}. Identifying ``Imagination'' in Kant's theory with the capability to select appropriate repertoires and his ``Understanding'' with the faculty of data compression of an aesthetic object or event, the ``free harmonious play'' of the mental faculties describes a never-ending, though converging process of aesthetic valuation, $\lim_{n \to \infty}\mathcal{X}_n$.

This interpretation is further supported by the neural reinforcement learning account suggested by \citet{Schmidhuber10a} and \citet{GoodfellowEA14}. As demonstrated by \citet{Graben24b}, Kant's ``Imagination'' can be identified with the creative generator module and ``Understanding'' with the categorizing discriminator module of a generative adversarial neural network (GAN) \citep{GoodfellowEA14, Graben24b}.

Finally, the GAN model provides a neurodynamical explanation of the experimental data from \citet{CheungHarrisonEA19}. In the neural network, Kant's ``free harmonious play'' is represented by an adversarial game between the generator module and the discriminator module, where the generator attempts to hocus the discriminator and vice versa \citep{Graben24b}. Then, high aesthetic pleasure for predictable closures in rather uncertain contexts reflects reward for the discriminator beating the generator, while high aesthetic pleasure for surprising closures in almost certain contexts indicates reward for the generator, this time beating the discriminator. This is consistent with phenomenological pragmatic information.


\section*{Acknowledgements}

I am gratefully indebted to Rubin Wang for his kind invitation to submit this work to \emph{Cognitive Neurodynamics}. Moreover, I thank Lea Schrader, Eduardo Mizraji and Reinhard Blutner for stimulating comments upon the manuscript. Computational resources provided by BTU Cottbus-Senftenberg through a guest scientist account are highly appreciated. Finally, I acknowledge financial support by the German Federal Agency of Labor.



\end{document}